\documentstyle[12pt]{article}

\catcode`\@=11
\long\def\@makefntext#1{
\protect\noindent \hbox to 3.2pt {\hskip-.9pt
$^{{\ninerm\@thefnmark}}$\hfil}#1\hfill}		%CAN BE USED

\def\@makefnmark{\hbox to 0pt{$^{\@thefnmark}$\hss}}  %ORIGINAL

\def\ps@myheadings{\let\@mkboth\@gobbletwo
\def\@oddhead{\hbox{}
\rightmark\hfil\ninerm\thepage}
\def\@oddfoot{}\def\@evenhead{\ninerm\thepage\hfil
\leftmark\hbox{}}\def\@evenfoot{}
\def\sectionmark##1{}\def\subsectionmark##1{}}

\renewcommand{\thefootnote}{\fnsymbol{footnote}}

\newcounter{sectionc}\newcounter{subsectionc}\newcounter{subsubsectionc}
\renewcommand{\section}[1] {\vspace*{0.6cm}\addtocounter{sectionc}{1}
\setcounter{subsectionc}{0}\setcounter{subsubsectionc}{0}\noindent
	{\normalsize\bf\thesectionc. #1}\par\vspace*{0.4cm}}
\renewcommand{\subsection}[1] {\vspace*{0.6cm}\addtocounter{subsectionc}{1}
	\setcounter{subsubsectionc}{0}\noindent
	{\normalsize\it\thesectionc.\thesubsectionc. #1}\par\vspace*{0.4cm}}
\renewcommand{\subsubsection}[1]
{\vspace*{0.6cm}\addtocounter{subsubsectionc}{1}
	\noindent {\normalsize\rm\thesectionc.\thesubsectionc.
\thesubsubsectionc.
	#1}\par\vspace*{0.4cm}}

\newcounter{appendixc}
\newcounter{subappendixc}[appendixc]
\newcounter{subsubappendixc}[subappendixc]

\renewcommand{\appendix}[1] {\vspace*{0.6cm}
        \refstepcounter{appendixc}
        \setcounter{figure}{0}
        \setcounter{table}{0}
        \setcounter{equation}{0}
        \renewcommand{\thefigure}{\Alph{appendixc}.\arabic{figure}}
        \renewcommand{\thetable}{\Alph{appendixc}.\arabic{table}}
        \renewcommand{\theappendixc}{\Alph{appendixc}}
        \renewcommand{\theequation}{\Alph{appendixc}.\arabic{equation}}
        \noindent{\bf Appendix \theappendixc #1}\par\vspace*{0.4cm}}

\def\abstracts#1{{
	\centering{\begin{minipage}{12.2truecm}\footnotesize
        \baselineskip=12pt\noindent
	\centerline{\footnotesize ABSTRACT}\vspace*{0.3cm}
	\parindent=0pt #1
	\end{minipage}}\par}}

\renewenvironment{thebibliography}[1]
	{\begin{list}{\arabic{enumi}.}
	{\usecounter{enumi}\setlength{\parsep}{0pt}
\setlength{\leftmargin 1.25cm}{\rightmargin 0pt}
	 \setlength{\itemsep}{0pt} \settowidth
	{\labelwidth}{#1.}\sloppy}}{\end{list}}

\newcounter{itemlistc}
\newcounter{romanlistc}
\newcounter{alphlistc}
\newcounter{arabiclistc}

\newcommand{\fcaption}[1]{
        \refstepcounter{figure}
        \setbox\@tempboxa = \hbox{\footnotesize Fig.~\thefigure. #1}
        \ifdim \wd\@tempboxa > 6in
           {\begin{center}
        \parbox{6in}{\footnotesize\baselineskip=12pt Fig.~\thefigure. #1}
            \end{center}}
        \else
             {\begin{center}
             {\footnotesize Fig.~\thefigure. #1}
              \end{center}}
        \fi}

\newcommand{\tcaption}[1]{
        \refstepcounter{table}
        \setbox\@tempboxa = \hbox{\footnotesize Table~\thetable. #1}
        \ifdim \wd\@tempboxa > 6in
           {\begin{center}
        \parbox{6in}{\footnotesize\baselineskip=12pt Table~\thetable. #1}
            \end{center}}
        \else
             {\begin{center}
             {\footnotesize Table~\thetable. #1}
              \end{center}}
        \fi}

\def\@citex[#1]#2{\if@filesw\immediate\write\@auxout
	{\string\citation{#2}}\fi
\def\@citea{}\@cite{\@for\@citeb:=#2\do
	{\@citea\def\@citea{,}\@ifundefined
	{b@\@citeb}{{\bf ?}\@warning
	{Citation `\@citeb' on page \thepage \space undefined}}
	{\csname b@\@citeb\endcsname}}}{#1}}

\newif\if@cghi
\def\cite{\@cghitrue\@ifnextchar [{\@tempswatrue
	\@citex}{\@tempswafalse\@citex[]}}
\def\citelow{\@cghifalse\@ifnextchar [{\@tempswatrue
	\@citex}{\@tempswafalse\@citex[]}}
\def\@cite#1#2{{$\null^{#1}$\if@tempswa\typeout
	{IJCGA warning: optional citation argument
	ignored: `#2'} \fi}}

 1
 1
 1

\font\ninerm=cmr9

\begin{document}

% define exp, dual fields,trace,plus sqrt,chi tilde,
%function of lambda or mu, cal A
%chi singular, chi new

\newcommand{\e}{\mbox{exp}}
\newcommand{\dual}{\stackrel{\circ}{\varphi}}
\newcommand{\p}{\stackrel{\circ}{p}}
\newcommand{\q}{\stackrel{\circ}{q}}
\newcommand{\tr}{\mbox{tr}}
\newcommand{\psqrt}[1]{_{+}\!\!\sqrt{#1}}
\newcommand{\sqrtI}[1]{_{{\rm I}}\!\sqrt{#1}}
\newcommand{\chit}{\tilde{\chi}}
\newcommand{\fl}{(\lambda)}
\newcommand{\fm}{(\mu)}
\newcommand{\A}{{\cal A}}
\newcommand{\chis}{\chi^{{\rm sing}}}
\newcommand{\Ct}{\tilde C}
\newcommand{\chin}{\chi^{{\rm new}}}
\newcommand{\Op}{{\cal O}}
\newcommand{\I}{\left(\begin{array}{cc}1&0\\0&1\end{array}\right)}

\centerline{\normalsize\bf PROBABILITY OF PHASE SEPARATION FOR THE}
\baselineskip=22pt
\centerline{\normalsize\bf BOSE GAS WITH DELTA INTERACTION}

\vfill
\vspace*{0.6cm}
\centerline{\footnotesize ALEXANDER R. ITS}
\baselineskip=13pt
\centerline{\footnotesize\it Department of Mathematical Sciences,
Indiana University-Purdue University at Indianapolis (IUPUI)}
\baselineskip=12pt
\centerline{\footnotesize\it Indianapolis, IN 46202-3216, U.S.A.}
\centerline{\footnotesize E-mail: itsa@math.iupui.edu}
\vspace*{0.3cm}
\centerline{\footnotesize and}
\vspace*{0.3cm}
\centerline{\footnotesize VLADIMIR E. KOREPIN}
\baselineskip=13pt
\centerline{\footnotesize\it Institute for Theoretical Physics,
State University of New York at Stony Brook}
\baselineskip=12pt
\centerline{\footnotesize\it Stony Brook, NY 11794-3840, U.S.A.}
\centerline{\footnotesize E-mail: korepin@max.physics.sunysb.edu}
\vspace*{0.3cm}
\centerline{\footnotesize and}
\vspace*{0.3cm}
\centerline{\footnotesize ANDREW K. WALDRON}
\baselineskip=13pt
\centerline{\footnotesize\it Institute for Theoretical Physics,
State University of New York at Stony Brook}
\baselineskip=12pt
\centerline{\footnotesize\it Stony Brook, NY 11794-3840, U.S.A.}
\centerline{\footnotesize E-mail: wally@max.physics.sunysb.edu}

\vfill
\vspace*{0.9cm}
\abstracts{We consider the quantum Non-linear Schr\"{o}dinger equation
$i\partial_t\Psi=-\partial_x^2\Psi +2c\Psi^\dagger\Psi^2$
with positive coupling constant $c$ varying from zero to infinity.
We study quantum correlation functions of this model using the determinant
 representation of these
correlation functions. We consider the case of a finite density
ground state and
evaluate, using the Riemann Hilbert problem,
the asymptotics of the probability
that no particles are present in the space
 interval $[0,x]$ in the large $x$ limit.
We call this the probability of phase separation or alternately
the emptiness formation probability.}

%\vspace*{0.6cm}
\normalsize\baselineskip=15pt
\setcounter{footnote}{0}
\renewcommand{\thefootnote}{\alph{footnote}}
\section{Introduction}
The one-dimensional Bose Gas is described by quantum Bose fields $\Psi(x,t)$
with canonical equal time commutation relations
$\left[\Psi(x),\Psi^\dagger(y)\right]=\delta(x-y)$,
The field $\Psi$ annihilates the Fock vacuum $\Psi|0\rangle=0$
and the  Hamiltonian of the model is
\begin{equation}
H=\int dx\left\{\partial_x \Psi^\dagger\partial_x\Psi+
c\Psi^\dagger\Psi^\dagger
\Psi\Psi\right\}.
\end{equation}
The equation of motion is the celebrated nonlinear Schr\"odinger equation
$i\partial_t\Psi=-\partial_x^2\Psi +2c\Psi^\dagger\Psi^2$.
The Hamiltonian and the number operator $Q=\int dx\Psi^\dagger\Psi$
commute whereby the model for a given number of particles $N$
is equivalent to the quantum mechanical Hamiltonian with delta
interactions
\begin{equation}{\cal H}_N=
           -\sum_{j=1}^{N}\frac{\partial^2}{\partial x^2_j}+2c
           \sum_{1\leq j<k\leq N}\delta(x_j-x_k).\end{equation}
The Bethe Ansatz for this
model was constructed by Lieb and Liniger~\cite{lieb}.

The ground state for the system in a box of length $L$ containing $N$
particles
in the thermodynamic limit ($N\rightarrow\infty$, $L\rightarrow\infty$,
$N/L=D=\mbox{const.}$) is a Fermi ``sphere'' of radius $q$. Introducing the
density of particles in momentum space $\rho(\lambda)$ so that
$D=\frac{N}{L}=\int_{-q}^{q}d\lambda\rho\fl$,
the ground state at zero temperature is then described by the linear integral
equation
\begin{equation}
\frac{1}{2\pi}=\rho\fl-\frac{1}{2\pi}\int_{-q}^{q}d\mu K(\lambda,\mu)\rho(\mu)
              =\rho\fl-\frac{1}{2\pi}\left(\hat{K}\rho\right)\!\fl,
\end{equation}
where the kernel of the integral operator $\hat{K}$ is
\begin{equation}
K(\lambda,\mu)=\frac{2c}{c^2+(\lambda-\mu)^2}.\label{K}
\end{equation}

Let us now lay out the structure of this paper.
We begin in  section 2 by defining the probability of
phase separation
correlation function $P(x)$ which we express in terms of the determinant
of a particular Fredholm integral operator. In section 3 we use the
identity $ic/(\lambda-\mu+ic)=c\int_{0}^{\infty}dse^{is(\lambda-\mu+ic)}$
to write this integral operator as a special Fredholm integral operator
related to a
Riemann--Hilbert problem (RP). The introduction of the $s$--integration
means that this RP is infinite-dimensional, or more precisely,
integral operator-valued.
In section 4 we present a new representation of $Gl(2,{\bf C})$
in terms of such integral operators which
reduces the jump condition~(\ref{bigjump})
of the operator RP to the jump condition~(\ref{smalljump})
of a $2\times2$ matrix-valued RP.
Finally in section 5 we solve the $2\times2$ RP for large $x$.
Indeed, we find that this $2\times 2$ RP is related to that encountered
if one studies the case $c=\infty$ for
which the $s$--integration is unnecessary.
The results obtained in section 5 lead us to a conjecture
(see \ref{conjecture})
for the asymptotic behaviour of the phase separation probability correlation
function. Presently we are unable to make this conjecture completely rigorous.

\section{Determinant Representation for Probability of Phase Separation
\label{efprob}}
Although the  ground state is homogeneous and translationally invariant, due
to quantum fluctuations there is a non-vanishing probability that
 no particles will be found in the space interval $[0,x]$ which we denote
\begin{equation}
P(x)=\mbox{Probability that no particles are found in the interval $[0,x]$}.
\end{equation}
In the thermodynamic limit the probability of an empty interval may be
expressed as the ratio of two Fredholm determinants~\cite{thebook}:
\begin{equation}
P(x)=\frac{(0|\det(I+\hat{V})|0)}{\det(I-\hat{K}/2\pi)}.\label{startpt}
\end{equation}
The integral operator $\hat{K}$ appearing in the denominator is given above
in~(\ref{K}) and is $x$ independent. Therefore when calculating the large
$x$ asymptotics we need consider only the numerator $(0|\det(I+\hat{V})|0)$.
The definition of the integral operator  $\hat{V}$ is more involved.
$\hat{V}$ acts on the interval $C=[-q,q]$ in the same
fashion as $\hat{K}$, namely
\begin{equation}
\left(\hat{V}f \right)\fl=\int_{-q}^{q}d\mu V(\lambda,\mu)\!f(\mu).
\end{equation}
However in addition to depending on the parameters $x$ and $c$
(the length of the empty interval and coupling respectively)
$\hat{V}$ and in turn $\det(I+\hat{V})$
are functionals of a dual quantum field $\dual\fl$. Let us give the
kernel of $\hat{V}$
\begin{eqnarray}
V(\lambda , \mu )& = &-\frac{c}{2\pi}
 \left[ \frac{\e\{ ix\lambda /2 +\dual (\lambda )/2 \} \e
\{ -ix\mu /2 -\dual (\mu )/2 \} }
			{(\lambda -\mu)(\lambda -\mu +ic)}\right.
					\nonumber\\
&& \hspace{1cm} \left. +\frac{\e \{ ix\mu /2 +\dual (\mu )/2\}
        \e \{ -ix\lambda /2 -\dual (\lambda )/2\}}
			{(\mu -\lambda)(\mu -\lambda +ic)}\right] .
\label{kernel}
\end{eqnarray}
The dual field acts in a Hilbert space with ``dual'' Fock vacuum $|0)$
so that
$(0|\det (I+\hat{V})|0)$ denotes the dual vacuum expectation value
of the dual field valued determinant $\det (I+\hat{V})$ which may be
computed from the
definition of the dual field $\dual\fl$ given below:
 \begin{equation}
\dual (\lambda )= \p (\lambda ) + \q (\lambda ),\label{dual1}
\end{equation}
where $\forall\; \lambda , \mu \in C$
\begin{equation}
[\p (\lambda ),\q (\mu )] =
\mbox{log}\left(\frac{c^2}{c^2+(\lambda-\mu )^2}\right), \end{equation}
\begin{equation} [\p (\lambda ),\p (\mu )] = 0
 =  [\q (\lambda ),\q (\mu )]\ ;\  \p (\lambda)|0) =   0
 = (0|\q (\lambda).\label{dual4}
\end{equation}
{}From the symmetry of $\mbox{log}(c^2/(c^2+(\lambda-\mu )^2))$ in $\lambda$
 and $\mu$
we have the crucial property
\begin{equation}[\dual (\lambda ),\dual (\mu )]=
0\ \ \ \ \  \forall \ \lambda, \mu \in C.\end{equation}
This allows us to treat $\dual \!(\lambda)$
simply as some function on $C$ in our calculation of $\det (I+\hat{V})$.
Only at the end of this calculation do we take the dual
vacuum expectation value.

Obtained using the Riemann Hilbert problem in the large $x$ limit, our
 conjecture for the
determinant as a functional of $\dual\fl$
is\footnote{$\psqrt{\lambda^2-q^2}$ denotes
the limit approaching the contour $C$ from above.
Further, we remind the reader that
the Hardy small ``$o$'' symbol $o(x)$ denotes any function of $x$
 decreasing faster than $x$
as $x\rightarrow \infty$.} :

\begin{equation}
\det(I+\hat{V})\stackrel{x\rightarrow\infty}{=}(\mbox{const})
\exp\left\{-(xq)^2/8 -\frac{x}{2\pi}\int_{-q}^q
\frac{\dual\fm\mu d\mu}{\psqrt{\mu^2 -q^2}}+o(x)\right\}.
\label{answer}
\end{equation}
Let us now analyse the dual vacuum expectation value of~(\ref{answer}).
Using~(\ref{dual1})-(\ref{dual4})
and the identity $e^{A+B}=e^Ae^Be^{-\frac{1}{2}[A,B]}$ valid $\forall$
 $A$, $B$ which commute with
$[A,B]$ we obtain
\begin{eqnarray}
(0|\det(I+\hat{V})|0)\!\!\!&\sim&\!\!\!\exp\left\{\!-\frac{(xq)^2}{8}
	-\frac{x^2}{8\pi^2}\int_{-q}^q\!\frac{\mu d\mu}{\psqrt{\mu^2-q^2}}
        \int_{-q}^q\!\frac{\nu d\nu}{\psqrt{\nu^2-q^2}}
	\log\frac{c^2\!+(\mu-\nu)^2}{c^2}\!\right\}\nonumber\\
	\!\!\!&\equiv&\!\!\!\exp\left\{\!-\frac{(xq)^2}{8}[1+I(c^2/q^2)]
\right\},
\end{eqnarray}
where we denote the dimensionless integral
\begin{equation}
I(c^2/q^2)=\frac{2}{\pi^2}\int_{-1}^1
\frac{ydy}{\sqrt{1-y^2}}\int_{-1}^1\frac{zdz}{\sqrt{1-z^2}}
		\log\left(\frac{c^2/q^2+(y+z)^2}{c^2/q^2+(y-z)^2}\right).
\end{equation}
$I(c^2/q^2)$ has the following properties:
$I(c^2/q^2) > 0$,
$\frac{\partial I}{\partial (c^2/q^2)} < 0$,
$I(c^2/q^2\rightarrow \infty)=0$ and
$I(c^2/q^2\rightarrow 0) =1$.
Therefore we find as our conjecture for the phase separation probability
 correlation function

\begin{equation}
P(x)\stackrel{x\rightarrow\infty}{=}
(\mbox{const})\exp\left\{-\frac{(xq)^2}{8}[1+I(c^2/q^2)]+o(x^2) \right\}.
\label{conjecture}
\end{equation}
The probablility of phase separation is Gaussian in the length $x$
and the rate of decay in $x$ decreases monotonically as the coupling
 $c$ varies from zero
to infinity.

\section{Integrable Linear Integral Operators and $s$--Integration}
\label{sint}
Integral operators $I+\hat{V}$ (acting on some interval $C$ say)
where the kernel of $\hat{V}$ has  the form
\begin{equation}
V(\lambda,\mu)=\frac{1}{\lambda-\mu}\sum_{j=1}^{N}e_j\fl E_j\fm,
 \label{factorkernel}
\end{equation}
with $0 =\sum_{j=1}^{N}e_j\fl E_j\fl$,
belong to a special class which we will call  ``integrable'' operators
(the functions $e_i\fl$ are $N$ linearly independent functions, continuous
and integrable on the interval $C$, similarly for $E_j$).
Indeed such integrable integral operators form a group (where multiplication
is defined in the usual way for integral operators) and are
closely related to a certain $N\times N$ matrix Riemann--Hilbert problem.

Therefore to  employ the RP for finite $c$ we must express the
kernel~(\ref{kernel})
 in the standard form above. To this end we rewrite
the second factor in the denominators  using an old identity made
 famous by Feynman
$\frac{1}{\lambda-\mu+ic}=-i\int_{0}^{\infty}dse^{is(\lambda-\mu+ic)}$.
The kernel~(\ref{kernel}) becomes
\begin{eqnarray}
V(\lambda,\mu)&=&\int_{0}^{\infty}
\frac{e_+ (\lambda|s) e_-(\mu|s) -e_-(\lambda|s) e_+ (\mu|s)}{\lambda-\mu}ds
		\\
	      &=&\int_{0}^{\infty}\frac{e_1 (\lambda|s) E_1(\mu|s) +
               e_2(\lambda|s) E_2 (\mu|s)}{\lambda-\mu}ds\label{opkernel}\\
\mbox{where }&&\! \!\!\!\!\!\!\!\!\!\!e_{\pm}(\lambda|s)=
		\sqrt{\frac{i}{2\pi}}\exp\{ \pm(ix\lambda /2 +
 \dual (\lambda )/2)\}
		\sqrt{c}\exp\{ \pm is\lambda -cs/2\},\\
\mbox{and }\ \! \ \ &&\!\!\!\!\!\!\!\!\!\!
(e_1 (\lambda|s) , e_2 (\lambda|s))=(e_+ (\lambda|s) , e_-(\lambda|s)),\\
&&\!\!\!\!\!\!\!\!\!\!(E_1 (\lambda|s) ,
E_2 (\lambda|s))=(e_- (\lambda|s) , -e_+ (\lambda|s)),
\end{eqnarray}
which is in the form~(\ref{factorkernel}) with the sum
$\sum_{j=1}^{N}$ generalized to
an integral $\int_{0}^{\infty}ds\sum_{j=1}^{2}$. We must,
therefore study a $2\times 2$
operator valued RP which we now define.

 The operator valued RP is to find the $2\times2$ matrix
 $\widehat{\chi}\fl$ whose entries are
integral operators acting on the interval $[0,\infty)$ and
functions of the complex plane ${\bf C}$
with kernels
\begin{equation}
\chi(\lambda|s,t)=\left(\begin{array}{cc}
			\chi_{11}(\lambda|s,t)&\chi_{12}(\lambda|s,t)\\&\\
			\chi_{21}(\lambda|s,t)&\chi_{22}(\lambda|s,t)
	\end{array}\right), \; \lambda\in{\bf C};s,t\in[0,\infty),
\end{equation}
such that:

\begin{enumerate}
\item $\chi(\lambda|s,t)$ is analytic $\forall$
 $\lambda \in {\bf C}\setminus C$.

\item $\chi (\lambda|s,t )$ is, however, not analytic
across the conjugation contour $C$,
rather the limits $\chi_+ (\lambda|s,t )$ and $\chi_- (\lambda|s,t )$,
approaching
$C$ from above and below respectively\footnote{We consider $C=[-q,q]$
as an oriented contour so that $\int_{C}\equiv \int_{-q}^{q}$.
 In general for any
oriented contour we will define ``$+$'' to be the limit from the
left side of the direction of travel.},
 satisfy the jump condition
\begin{equation}
\chi_-^{ik}(\lambda|s,t)=
\int_{0}^{\infty}\sum_{j=1}^{2}\chi_+^{ij}
(\lambda|s,r)G^{jk}(\lambda|r,t)dr,\;
\ \forall \ \lambda \in C, \ (i,k=1,2),\label{bigjump}
\end{equation}
where the conjugation matrix $G_{ik}(\lambda|s,t)$ is given by
\begin{equation}
G_{ik}(\lambda|s,t)=\delta_{ik}\delta(s-t)-2\pi ie_i (\lambda|s)
 E_k (\lambda|t), \; \ (i,k=1,2).\label{opconjmtx}\label{ernie}
\end{equation}

\item $\chi(\lambda|s,t)$ is canonically normalized:
\begin{equation}
\chi (\infty |s,t)= \I\delta(s-t).
\end{equation}
In fact, if  $\widehat{\chi}\fl$ solves the RP defined above,
 then at $\lambda=\infty$ it has the
whole Laurent series expansion $\chi(\lambda|s,t)=\I\delta(s-t)+
 \sum_{n=1}^{\infty}\frac{M_n(s,t)}{\lambda^n}$
where the  integral operators $\widehat{M}_n$ are
independent of $\lambda$.
\end{enumerate}

In order to extract the determinant $\det(I+\hat{V})$
from the operator RP we consider the
formal solution to the above RP
\begin{eqnarray}
\chi _{ik}(\lambda|s,t)&=&\delta_{ik}\delta(s-t)
	+\int_{C}\frac{f_i (\mu|s )
 E_k (\mu|t )d\mu}{\mu-\lambda}\label{opformalsoln}\\
\mbox{where }\ \ \ \ &&\!\!\!\!\!\!\!\!\!\!\!\!\!\!\!\!\!\!\!\!e_i
 (\lambda|s)= [(I+\hat{V})f_i](\lambda|s)=f_i (\lambda|s)+\int_{C}
V(\lambda,\mu ) f_i (\mu|s)d\mu .
\end{eqnarray}
For $\lambda$ large we have
\begin{eqnarray}
\chi_{ik}(\lambda|s,t)&\stackrel{\lambda \rightarrow \infty}{\sim}
&\delta_{ik}\delta(s-t)-
\frac{1}{\lambda}\int_{C}f_i(\mu|s)
 E_k (\mu|t)d\mu+ \mbox{O}(\lambda^{-2})\\
&=&\delta_{ik}\delta(s-t)+\frac{M_1^{ik}(s,t)}{\lambda}+
 \mbox{O}(\lambda^{-2}).
\end{eqnarray}
But now (see~\cite{thebook}) calculate the
 logarithmic derivative w.r.t. $x$ of $\det(I+\hat{V})$ from~(\ref{opkernel})
\begin{eqnarray}
\frac{\partial}{\partial x}\log\det
 (I+\hat{V})&=&\frac{i}{2}\int_{C}\int_{0}^{\infty}
(f_1(\mu|s) e_2(\mu|s)+f_2(\mu|s) e_1(\mu|s))dsd\mu \nonumber\\
  &=&-\frac{i}{2}\int_{0}^{\infty}(M_1^{11}(s,s)-M_1^{22}(s,s))ds
			      =-i\mbox{tr}\widehat{M}_1^{11},\label{opdet}
\end{eqnarray}
where ``tr'' denotes the trace operation in the space of
 integral operators on $[0,\infty)$.
Therefore $\det(I+\hat{V})$ can be obtained from the large
 $\lambda$ asymptotics of the
solution to the above RP.

At this point the prospect of solving an \underline{operator}
valued RP should seem daunting since even the solution for the
matrix  case is not in general known.
However we now construct a representation of $Gl(2,{\bf C})$ which will allow
us to consider a (known) $2\times 2$ RP.

\section{$Gl(2,{\bf C})$ Represention by Integral Operators}
Let us consider some $2\times 2$ matrix of integral operators
$\widehat{\Op}$ with kernels

\begin{equation}
\Op(s,t)=\left(\begin{array}{cc}
			\Op_{11}(s,t)&\Op_{12}(s,t)\\&\\
			\Op_{21}(s,t)&\Op_{22}(s,t)
			\end{array}\right), \; s,t\in[0,\infty),
\end{equation}
and  multiplication defined by\footnote{We assume that the
 integral operators $\widehat{\Op}_{ik}$
act on the interval $[0,\infty)$ but this is of course not necessary.}
\begin{equation}
\Op_{{\rm III}}(s,t)=
\left(\widehat{\Op}_{{\rm I}}\widehat{\Op}_{{\rm II}}\right)(s,t)
=\left(\begin{array}{cc}
\Op^{{\rm III}}_{11}(s,t)&\Op_{12}^{{\rm III}}(s,t)\\&\\
			\Op_{21}^{{\rm III}}(s,t)&\Op_{22}^{{\rm III}}(s,t)
			\end{array}\right), \; s,t\in[0,\infty),
\end{equation}
where
\begin{eqnarray}
\Op^{{\rm III}}_{11}(s,t)&=&
\int_{0}^{\infty}dr\left(\Op^{{\rm I}}_{11}(s,r)\Op^{{\rm II}}_{11}(r,t)
      +\Op^{{\rm I}}_{12}(s,r)\Op^{{\rm II}}_{21}(r,t)\right)\nonumber \\
\Op^{{\rm III}}_{12}(s,t)&=&
\int_{0}^{\infty}dr\left(\Op^{{\rm I}}_{11}(s,r)\Op^{{\rm II}}_{12}(r,t)
         +\Op^{{\rm I}}_{12}(s,r)\Op^{{\rm II}}_{22}(r,t)\right)\nonumber \\
\Op^{{\rm III}}_{21}(s,t)&=&
\int_{0}^{\infty}dr\left(\Op^{{\rm I}}_{21}(s,r)\Op^{{\rm II}}_{11}(r,t)
       +\Op^{{\rm I}}_{22}(s,r)\Op^{{\rm II}}_{21}(r,t)\right)\nonumber \\
\Op^{{\rm III}}_{22}(s,t)&=&
\int_{0}^{\infty}dr\left(\Op^{{\rm I}}_{21}(s,r)\Op^{{\rm II}}_{12}(r,t)
             +\Op^{{\rm I}}_{22}(s,r)\Op^{{\rm II}}_{22}(r,t)\right).
\end{eqnarray}
Let us now construct a special class of such operators
 $\widehat{\Op}$ which form a representation
of  $Gl(2,{\bf C})$ (the generalization to $Gl(N,{\bf C})$
 is straightforward).
Consider a pair of functions $\alpha(s)$ and $\beta(s)$
 on $[0,\infty)$ which we write
in Dirac notation as $\alpha(s)\equiv|1\rangle$
and $\beta(s)\equiv|2\rangle$.
In this notation we may write left multiplication by $\widehat{\Op}_{ik}$ as
\begin{equation}
\widehat{\Op}_{ik}|1\rangle=\int_{0}^{\infty}dt\Op_{ik}(s,t)\alpha(t).
\end{equation}
Further suppose the functions
$A(s)\equiv\langle 1|$ and $B(s)\equiv\langle 2|$ (right multiplication by
$\widehat{\Op}_{ik}$ is defined integrating over
 the first argument of $\Op_{ik}(s,t)$) satisfy
\begin{equation}
\langle1|1\rangle\equiv\int_{0}^{\infty}dsA(s)\alpha(s)=1=
\langle2|2\rangle\equiv\int_{0}^{\infty}dsB(s)\beta(s).
\end{equation}
Observe now that one may define a representation
 $\widehat{\cal{A}}$ of $Gl(2,{\bf C})$
via
\begin{equation}
M\!\in Gl(2,{\bf C})\longmapsto{\widehat{\cal A}}(M)=\left(\begin{array}{cc}
					 I-|1\rangle\langle1| & 0\\&\\
					 0 & I-|2\rangle\langle2|
                                         \end{array}\right)
				+	 \left(\begin{array}{rr}
	 M_{11}|1\rangle\langle1| & M_{12}|1\rangle\langle2|\\&\\
			 M_{21}|2\rangle\langle1| & M_{22}|2\rangle\langle2|
                                         \end{array}\right),
\end{equation} ($M_{11}$, $M_{12}$, $M_{21}$  and $M_{22}$
 are complex numbers and
$I$ is the identity operator in the space of integral oprators on
$[0,\infty)$).
Multiplication by the integral operators (projectors)
$|1\rangle\langle1|$, $|1\rangle\langle2|$, $|2\rangle\langle1|$ and
$|2\rangle\langle2|$
is given, for example, by
\begin{equation}
|1\rangle\langle2|f(s)=\left(\int_{0}^{\infty}dsB(s)f(s)\right)|1\rangle.
\end{equation}
In particular, we have
$\left[ I-|1\rangle\langle1|\right] |1\rangle\langle1|=0$.
Indeed for any $M$, $N$ $\in$ $Gl(2,{\bf C})$
the representation $\widehat{\cal A}$ has the following properties:
\begin{eqnarray}
\lefteqn{\widehat{\cal A}(MN)=\widehat{\cal A}(M)\widehat{\cal A}(N)\ ;
\ \widehat{\cal A}(I)=I \ ;\ \widehat{\cal A}(M^{-1})=
\widehat{\cal A}^{-1}(M)}\label{repn}\\
\lefteqn{\det\widehat{\cal A}(M)=\det M=
M_{11}M_{22}-M_{12}M_{21}}\label{detA}\\
\lefteqn{\tr\left(\widehat{\cal A}(M)-\left(\begin{array}{cc}
					 I-|1\rangle\langle1| & 0\\&\\
					 0 & I-|2\rangle\langle2|
        \end{array}\right)\right)=\tr M=M_{11}+M_{22}}&&\hspace{8.5cm}
\end{eqnarray}
Notice that~(\ref{detA}) expresses an infinite dimensional determinant
as the determinant
of the $2\times2$ matrix $M$.
Let us conclude this section by rewriting the conjugation matrix of our RP
using the representation $\widehat{\cal A}$.
We begin by defining functions
\begin{equation}
\left\{ \begin{array}{l}
       |1\rangle=\sqrt{c}e^{is\lambda -cs/2}=\alpha(s), \;  \;\langle1|
=\sqrt{c}e^{-is\lambda -cs/2}=A(s)\\ \\
       |2\rangle=\sqrt{c}e^{-is\lambda -cs/2}=\beta(s), \; \;\langle2|
=\sqrt{c}e^{is\lambda -cs/2}=B(s)
       \end{array}\right.
\end{equation}
The vectors $|1\rangle$ and $|2\rangle$ are
 normalized but not orthogonal, so
that their inner products are
$\langle1|1\rangle=c\int_{0}^{\infty}e^{-cs}ds=1=\langle2|2\rangle$ and
$\langle1|2\rangle=\langle2|1\rangle^*=
c\int_{0}^{\infty}e^{-2is\lambda-cs}ds=\frac{c}{c+2i\lambda}$.
(for $\lambda$ real the bra and ket notation denote complex
conjugation in the
usual fashion although this is not necessary).
Their outer products are of course integral operators  in terms of which
we will write the conjugation matrix.
We now have
\begin{equation}
\left\{ \begin{array}{l}
       e_1(\lambda|s)=\sqrt{\frac{i}{2\pi}}e^{(ix\lambda /2 +
 \dual (\lambda )/2)}|1\rangle
\equiv e^{\circ}_1\fl |1\rangle,\\
       E_1(\lambda|t)=\sqrt{\frac{i}{2\pi}}e^{(-ix\lambda /2 -
\dual (\lambda )/2)}\langle1|
\equiv E^{\circ}_1\fl \langle1|,\\
       e_2(\lambda|s)=\sqrt{\frac{i}{2\pi}}e^{(-ix\lambda /2 -
\dual (\lambda )/2)}|2\rangle
\equiv e^{\circ}_2\fl |2\rangle,\\
       E_2(\lambda|t)=-\sqrt{\frac{i}{2\pi}}e^{(ix\lambda /2 +
\dual (\lambda )/2)}\langle2|
\equiv E^{\circ}_2\fl \langle2|.
	     \end{array}\right.
\end{equation}
Hence, from~(\ref{ernie}) we have for the conjugation matrix (where from
now on the $s$ and $t$ dependence will be taken as understood)
\begin{equation}
\widehat{G}\fl=\left(\begin{array}{rr}
           I-2\pi ie^{\circ}_1\fl E^{\circ}_1\fl|1\rangle\langle1| &
-2\pi ie^{\circ}_1\fl E^{\circ}_2\fl|1\rangle\langle2|\\&\\
           -2\pi ie^{\circ}_2\fl E^{\circ}_1\fl|2\rangle\langle1|
 &I-2\pi ie^{\circ}_2\fl E^{\circ}_2\fl|2\rangle\langle2|
           \end{array} \right)={\widehat{\cal A}}(G^{\circ}\fl).
\label{jmpmtrx}
\end{equation}
where the $2\times 2$ matrix $G^{\circ}\fl$ which here
 plays the r\^{o}le of $M$ is readily computed to be
\begin{equation}G^\circ(\lambda )=\left(\begin{array}{cc} 2
 & -e^{ix\lambda+\dual (\lambda )}\\&\\
  e^{-ix\lambda-\dual (\lambda )}& 0\end{array}\right).\label{elvis}
\end{equation}
Except for the additional exponents $\pm\dual\fl$, $G^{\circ}\fl$ is the
conjugation matrix one finds for the $2\times2$ RP encountered
 in the case $c=\infty$.

\section{Asymptotic Analysis of the Riemann Hilbert Problem}
Firstly let us sketch the derivation of the asymptotic solution to the
$2\times2$ RP with conjugation matrix $G^\circ\fl$ given above
in~(\ref{elvis}) (we  follow the scheme developed in~\cite{its} for
 the case $\dual\fl\equiv 0$).
I.e. we are searching for the $2\times2$ matrix function
$\chi^\circ\fl$ normalized to $\I$ at infinity and analytic
everywhere in the
complex plane ${\bf C}$ save for a discontinuity across the
 contour $C=[-q,q]$
where  $\chi^\circ\fl$ satisfies
\begin{equation}
\chi^\circ_-\fl=\chi_+^\circ\fl G^\circ\fl,\ \
\forall \lambda \in C.\label{smalljump}
\end{equation}
The idea is to make a ``gauge transformation'' under which
the above RP limits for large $x$  to a new RP with conjugation
matrix $\tilde{G}=\left(\begin{array}{cc}0&-1\\1&0\end{array}\right)$.
Define, therefore
\begin{equation}
\chit^\circ\fl=\exp\left\{\psi_\infty\sigma_3\right\}\chi^\circ\fl
\exp\left\{[(ix/2)(\lambda-g\fl)-\psi\fl]\sigma_3\right\},
\label{trans1}
\end{equation} ($\sigma_3=\mbox{diag}(1,-1)$).
The function $g\fl=\sqrt{\lambda^2-q^2}$ satisfies the scalar RP
 $g_-\fl=-g_+\fl$ across
$C$ with normalization
 $g\fl\stackrel{\lambda\rightarrow\infty}{\sim}
\lambda-q^2/2\lambda +O(\lambda^{-3})$
 and
 \begin{equation}
\psi(\lambda )=-\frac{\sqrt{\lambda^2 -q^2}}{2\pi i}
\int_{C}\frac{\dual (\mu)d\mu}{\psqrt{\mu^2 -q^2}(\mu-\lambda)}\ .
\end{equation}
satisfies the  affine scalar RP $\psi_+ (\lambda )
 +\psi_- (\lambda) +\dual (\lambda )=0$
across the contour $C$ with normalization $\psi(\lambda)
 \stackrel{\lambda\rightarrow
\infty}{\longrightarrow} \psi_\infty =\frac{1}{2\pi i}
\int_C \frac{\dual (\mu)d\mu}{\psqrt{\mu^2 -q^2}}$.
Notice that $\psi\fl$ is well defined on all of ${\bf C}$
 even though it depends on the dual
quantum field $\dual\fl$.
One may verify that $\chit\fl$ satisfies a canonically
 normalized RP across $C$
with conjugation matrix
\begin{eqnarray}
\tilde{G}^\circ\fl \!\!\!\!&=&\!\!\! \exp\!
\left\{\!-[(ix/2)(\lambda\!-g_+\fl)\!-\!\psi_+\!\fl]
\sigma_3\right\}\!G^\circ\fl
    \exp \!\left\{[(ix/2)(\lambda\!-\!g_-\fl)\!-\!\psi_-\fl]
\sigma_3\right\}\nonumber\\
&& \nonumber\\
          \!\!\!\!  &=&\!\!\!
\left( \begin{array}{cc}2e^{\Delta\psi(\lambda)+ixg_+\fl}&-1 \\&\\
				      1&0\end{array}\right)
\stackrel{x\rightarrow \infty}{\longrightarrow}\left(
 \begin{array}{cc}0&-1 \\&\\
				      1&0\end{array}\right),
\end{eqnarray}
where $\Delta\psi\fl=\psi_+\fl-\psi_-\fl$.
The canonically normalized RP for $\chit_\fl$
is easily solved in the large $x$ limit,  we find
\begin{equation}
\chit^\circ\fl\stackrel{x\rightarrow\infty}{\sim}\left(
 \begin{array}{cc}\frac{1}{2}(a\fl + a^{-1}\fl)&
\frac{i}{2}(a\fl -a^{-1}\fl) \\&\\
				    -\frac{i}{2}(a\fl -a^{-1}\fl)
 &\frac{1}{2}(a\fl + a^{-1}\fl)\end{array}\right),
\end{equation}
where $a\fl=(\frac{\lambda+q}{\lambda-q})^{1/4}$ satisfies the canonically
normalized RP $a_-\fl=ia_+\fl$ ($\forall \lambda \in C$).
Inverting the transformation~(\ref{trans1}) we obtain the large $x$
solution to the original RP,
\begin{eqnarray}
\chi^\circ\fl\!\!\!&\stackrel{x\rightarrow\infty}{\sim}&\!\!\!\!\left(\!\!
\begin{array}{cc}\frac{1}{2}\!(\!a\fl \!+ \!a^{-1}\!\fl\!)
e^{\psi\!\fl\!-\!\psi_\infty\! -\!ix(\lambda\!-\!g\!\fl)/2}
		 &\frac{i}{2}\!(\!a\fl \!-\!a^{-1}\!\fl\!)
e^{-\psi\!\fl\!-\!\psi_\infty\! +\!ix(\lambda\!-\!g\!\fl)/2}\\&\\
		 -\frac{i}{2}\!(\!a\fl \!-\!a^{-1}\!\fl\!)
 e^{\psi\!\fl\!+\!\psi_\infty\! -\!ix(\lambda\!-\!g\!\fl)/2}
		&\frac{1}{2}\!(\!a\fl \!+\! a^{-1}\!\fl\!)
e^{-\psi\!\fl\!+\!\psi_\infty\!+\!ix(\lambda\!-\!g\!\fl)/2}
\end{array}\!\!\right)\nonumber
		\label{2times2soln}\\ && \nonumber \\
\!\!\!&=&\!\!
I+\frac{1}{\lambda}\left(
\begin{array}{cc}\psi_1 -ixq^2/4
		 &\frac{iq}{2}e^{-2\psi_\infty}\\&\\
		 -\frac{iq}{2}e^{2\psi_\infty}
		&-\psi_1 +ixq^2/4\end{array}\right)+\mbox{O}
(\lambda^{-2}),\label{solnexpansion}
\end{eqnarray}
where $\psi_1=
\frac{1}{2\pi i}\int_C \frac{\dual (\mu)\mu d\mu}{\psqrt{\mu^2 -q^2}}$.

We may now try to utilise our $Gl(2,{\bf C})$ representation to
 write the solution
to the finite $c$ operator valued RP as
\begin{equation}
\widehat{\chi}\fl=\widehat{\cal A}(\chi^\circ\fl),\label{userep}
\end{equation}
since by property~(\ref{repn})
\begin{equation}
\widehat{\chi}_-\fl=\widehat{\cal A}(\chi^\circ_-\fl)=
\widehat{\cal A}(\chi^\circ_+\fl G^\circ\fl)
=\widehat{\cal A}(\chi^\circ_+\fl )\widehat{\cal A}(G^\circ\fl)=
\widehat{\chi}_+\fl \widehat{G}\fl.
\end{equation}
The observant reader may notice that this solution does not obey the
normalization at infinity since the projectors
$|1\rangle\langle 1|,|1\rangle\langle 2|,|2\rangle\langle 1|$ and
$|2\rangle\langle 2|$
are essentially singular at $\lambda=\infty$. Moreover,
 as bounded operator valued functions
they are analytic in the strip $|\mbox{Im}\lambda|<c/2$ only.
Therefore strictly speaking equation~(\ref{userep})  is not correct.
The correct version is
\begin{equation}
\widehat{\chi}\fl=\widehat{\Phi}\fl\widehat{\cal A}(\chi^\circ\fl),
\end{equation}
where $\widehat{\Phi}\fl$ is an operator valued function analytic in the strip
${\cal S}:|\mbox{Im}\lambda|<c/2$. The last equation itself may be treated
as a Riemann Hilbert Problem. Indeed, let $\Gamma$ be a simple contour
belonging
to the strip ${\cal S}$ encircling the interval $C$ in the counterclockwise
direction. Introducing
\begin{equation}
\widehat{Y}\fl=\left\{ \begin{array}{l}\widehat{\chi}\fl
 \mbox{ if $\lambda$ is outside $\Gamma$} \\
                             \widehat{\Phi}\fl \mbox{ if $\lambda$
is inside $\Gamma$,}\end{array}\right.
\end{equation}
we can rewrite our initial operator RP 1-3 as the following new
RP\footnote{
To avoid confusion, in agreement with our conventions let us
 reiterate that $+$ is
 the limit from \underline{inside} the anticlockwise contour $\Gamma$.} :
\begin{description}
\item[$\ \ \ \  1^\Gamma.$] $\widehat{Y}\fl$ is analytic
$\forall \lambda \in {\bf C}\setminus \Gamma$.

\item[$\ \ \ \  2^\Gamma.$]
 $\widehat{Y}_-\fl=\widehat{Y}_+\fl\widehat{\cal A}(\chi^\circ\fl)$,
 $\lambda \in \Gamma$.

\item[$\ \ \ \  3^\Gamma.$] $\widehat{Y}(\infty) =I\ \I.$
\end{description}
{}From~(\ref{opdet}) and the equation
\begin{equation}
\det \widehat{\chi}_{11}\fl=1+
 \tr \widehat{M}_1^{11}/\lambda  +O(\lambda^{-2}),
\end{equation}
we see that to calculate $\det(I+\hat{V})$ we need $\det \widehat{Y}^{11}\fl$
for large $\lambda$.
It follows from 2$^\Gamma$ that
\begin{equation}
\det \widehat{Y}^{11}_+\fl=\det
\widehat{Y}^{11}_-\fl\left( \chi^\circ_{22}\fl
 -\chi^\circ_{21}\fl c\fl\right),\label{srp}
\end{equation}
where ``$\det$'' indicates the determinant of integral operators
 on the space $[0,\infty)$.
and $c\fl=\langle1|[\widehat{Y}^{11}_-\fl]^{-1}\widehat{Y}^{12}_-\fl|2\rangle
=\langle1|[\widehat{\chi}^{11}\fl]^{-1}\widehat{\chi}^{12}\fl|2\rangle$.
Let us now \underline{assume} that
\begin{description}
\item[$\ \ \ \  {\rm (a)}$] $c\fl$ is analytic outside $\Gamma$,
\item[$\ \ \ \  {\rm(b)}$] $c\fl =O(1/\lambda)$ as
$\lambda\rightarrow \infty$.
\end{description}
Note that the product $[\widehat{Y}^{11}_-\fl]^{-1}\widehat{Y}^{12}_-\fl$
($=[\widehat{\chi}^{11}_-\fl]^{-1}\widehat{\chi}^{12}_-\fl$)
has these properties.

Statements (a) and (b) constitute our main hypothesis concerning
 the solution $\chi\fl$ of the RP 1-3.
By virtue of this hypothesis, equation~(\ref{srp}) implies that
\begin{equation}
\det \widehat{Y}^{11}\fl=(\chi^\circ_{22}\fl-\chi^\circ_{21}\fl c\fl)^{-1},
\end{equation}
for all $\lambda$ outside of the contour $\Gamma$. In addition property (b)
leads to the equation
\begin{equation}
\lim_{\lambda\rightarrow\infty} \lambda(\det \widehat{Y}^{11}\fl-1)=
-\lim_{\lambda\rightarrow\infty} \lambda(\chi^\circ_{22}\fl-1).
\end{equation}
Taking~(\ref{srp}) and~(\ref{solnexpansion}) into account we obtain
\begin{equation}
\tr \widehat{M}^{11}_1\stackrel{x\rightarrow\infty}{=}\psi_1-ixq^2/4+o(1),
\end{equation}
which yields our conjecture for the determinant
\begin{equation}
\det(I+\hat{V})\stackrel{x\rightarrow\infty}{=}
(\mbox{const})\exp\left\{-(xq)^2/8
      -\frac{x}{2\pi}\int_C
 \frac{\dual\fm\mu d\mu}{\psqrt{\mu^2 -q^2}}+o(x)\right\},
\end{equation}
as quoted in section 2.
Finally let us observe that for $c=\infty$, $[\p (\lambda ),\q (\mu )]=0$
 so that
for any functional $F[\dual (\lambda )]$ the dual vacuum expectation value
$(0|F[\dual (\lambda )]|0)=F\left| _{\dual =0}\right.$.
Hence
$(0|\det(I+\hat{V})|0)\left|_{c=\infty}\right.=e^{-(xq)^2/8}$.
An old result first obtained in~\cite{oiseaux} and then extended to
 the complete asymptotic
expansion in~\cite{dyson}.

\section{Acknowledgements}
This work was supported by NSF grant PHY-9321165.

\end{document}